\documentclass[12pt,epsfig]{article}
\usepackage{graphicx}%

\newcommand{\eq}[1]{\begin{equation} #1 \end{equation}}
\newcommand{\ar}[1]{\begin{eqnarray} #1 \end{eqnarray}}
\newcommand{\tr}{\mathop{\mathrm{tr}}\nolimits}

\def\e{{\,\rm e}\,}
\def\d{\partial}

\def\dd{^{\dagger}}
\newcommand{\br}[1]{\left( #1 \right)}
\newcommand{\vev}[1]{\left\langle #1 \right\rangle}
\newcommand{\rf}[1]{(\ref{#1})}
\newcommand{\non}{\nonumber \\*}
\renewcommand{\/}{\!\!\!/}

\def\htc{high-$T_c$~}
\def\o{\omega}

\def\appendix#1{
  \addtocounter{section}{1}
 \setcounter{equation}{0}
  \renewcommand{\thesection}{\Alph{section}}
 \section*{Appendix \thesection\protect\indent \parbox[t]{11.715cm} {#1}}
  \addcontentsline{toc}{section}{Appendix \thesection\ \ \ #1}
  }

\begin{document}
\begin{titlepage}
\begin{flushright}
ITEP--TH--17/01\\
\end{flushright}
\vspace{1.5cm}

\begin{center}
{\LARGE Possible Pseudogap Phase in QCD} \\
\vspace{1.9cm}
{\large K.~Zarembo}\\
\vspace{24pt}
{\it Department of Physics and Astronomy}
\\{\it and Pacific Institute for the Mathematical Sciences}
\\{\it University of British Columbia}
\\ {\it 6224 Agricultural Road, Vancouver, B.C. Canada V6T 1Z1} 
\\ \vskip .2 cm
and\\ \vskip .2cm
{\it Institute of Theoretical and Experimental Physics,}
\\ {\it B. Cheremushkinskaya 25, 117259 Moscow, Russia} \\ \vskip .5 cm
E-mail: {\tt zarembo@physics.ubc.ca}
\end{center}
\vskip 2 cm
\begin{abstract}          
Thermal pion fluctuations, in principle, can completely disorder
the phase of the quark condensate and thus restore chiral symmetry.
If this happens before the quark condensate melts, strongly-interacting matter
will be in the pseudogap state just above the chiral phase transition.
The quark condensate does not vanish locally and 
quarks acquire constituent masses in the pseudogap phase, despite
chiral symmetry is restored. 
\end{abstract}
\end{titlepage}

The physics of light hadrons is to a large extent controlled by
the approximate chiral symmetry of QCD. The order parameter
of the chiral symmetry, the quark condensate $\vev{\bar{\psi}\psi}$, 
acquires a non-zero 
expectation value in the QCD vacuum and chiral symmetry
appears spontaneously broken. 
Chiral symmetry breaking  gives quarks  constituent
masses of order of $350$-$400$~Mev and thus sets the scale of hadron masses.
Pions
arise as low-energy, pseudo-Goldstone excitations of the chiral condensate.
If strongly-interacting matter is heated to the temperature 
that exceeds a critical value of  $150$-$200$~Mev, 
chiral symmetry gets restored. 
Usually, the restoration of  chiral symmetry is associated with melting of the quark
condensate.
I will discuss a less familiar mechanism of symmetry restoration by phase
decoherence which, if realized, implies the existence of 
an intermediate phase, 
similar to the pseudogap
phase of \htc superconductors,  the analogy with which I will extensively use. 
In the pseudogap phase, quarks still condense and
acquire constituent masses, 
but chiral symmetry is not broken because the phase of the condensate is completely
disordered.  
The potential relevance of the pseudogap phenomenon to QCD
was pointed out by Babaev and Kleinert \cite{Kle98,Bab01}, 
who examined low-dimensional toy models of chiral symmetry breaking
\cite{Kle98,Bab01,Mac94} and Nambu-Jona-Lasinio model \cite{Bab01, Bab00}.


A continuous symmetry associated with the complex order parameter
$\Psi=\rho\e^{i\varphi}$ is usually restored when the free energy of the
symmetry-breaking state with non-vanishing condensate $\rho\neq 0$  
starts to exceed the free energy of the symmetric
state with $\rho=0$. However, phase
decoherence can restore the symmetry even if $\rho\neq 0$.
When the phase of the condensate is completely disordered:  
$\vev{\e^{i\varphi}}=0$,
the expectation value of the  order parameter, 
apparently, vanishes: $\vev{\Psi}=0$. 
There is a growing evidence that such a mechanism 
is realized
in some \htc superconductors \cite{hightctheor,hightcexp}, whose
normal, non-superconducting state
possesses many features characteristic of
superconductivity.  Most notably, the energy gap does not
close above the point at which superconductivity is destroyed
and gradually disappears only at much higher temperature. A 
theoretical explanation of the pseudogap phenomenon \cite{hightctheor}
relies on the essentially
two-dimensional nature of \htc 
superconductivity. In two dimensions, topological phase fluctuations of the superconducting
condensate (vortices) are extremely important. Depending on the temperature and on 
the phase stiffness (the energy cost of phase fluctuations), vortices 
either are bound in pairs or form a plasma. 
These two phases are separated by Berezinsky-Kosterlitz-Thouless (BKT) transition
\cite{Ber71}.
In \htc superconductors, the temperature of BKT transition is lower than 
the temperature at which
the condensate of Cooper pairs melts and superconducting gap
accordingly shrinks. The superconductivity is then 
destroyed by vortices that unbind at the BKT transition
and completely disorder the condensate's phase.

In QCD, the chiral order parameter is  an $N_f\times N_f$ matrix
($N_f$ is the number of light quark species):
 $\Sigma_{ff'}=\bar{\psi}_f
\psi_{f'}$. 
Its vacuum expectation value is diagonal 
because light quarks have small current masses
which align the condensate in the particular direction.
Throughout this paper I will discuss the chiral limit and neglect quark masses.
The coherent fluctuations of the
condensate's phase cost no energy in this approximation
and correspond to Goldstone modes
of the broken chiral symmetry. The chiral phase of the quark condensate
 is an unitary $N_f\times N_f$ matrix:
$$
\e^{i\gamma^5\lambda^a\pi^a}=
\frac{1-\gamma^5}{2}\,U+\frac{1+\gamma^5}{2}\,U\dd\,,
$$
where
 $\lambda^a$ are $SU(N_f)$ generators 
and $\pi^a$ are Goldstone fields associated with pions.

The low-energy dynamics of pions is described by the chiral
Lagrangian:
\eq{\label{nlsm}
{\cal L}_\chi
=\frac{F^2_\pi}{4}\,\tr \d_\mu U\dd\d^\mu U, 
}
where $F_\pi=93$~Mev is the pion decay constant.
The classical thermodynamics of the
non-linear sigma model has been extensively studied
by Monte Carlo simulations \cite{Kog82,DeT90} and, indeed, the
 phase transition associated with disordering of the chiral field
has been found in numerical simulations:
$\vev{\tr U}\neq 0$ below the critical point and $\vev{\tr U}= 0$ above.
The  large-$N$ analysis \cite{Boc96} suggests that the phase transition
does not disappear if one goes from classical thermodynamics to quantum.
Since
\eq{
\vev{\bar{\psi}\psi}=\vev{|\Sigma|(\tr U+\tr U\dd)},}
the chiral condensate turns to zero 
when $\vev{\tr U}= 0$ 
and chiral symmetry gets restored.
The question is whether the temperature of pion disordering
is lower or higher
than the temperature at which the quark condensate melts.
If it is lower, as suggested by the smallness of the pion decay constant
(the analog of phase stiffness in QCD), then the pseudogap phase
exists in the intermediate range of temperatures. The 
results of Refs.~\cite{DeT92,Boc96} seem to
support  the hypothesis that 
the chiral phase transition is essentially driven by angular
fluctuations of the quark condensate.

\begin{figure}
\begin{center}
\includegraphics{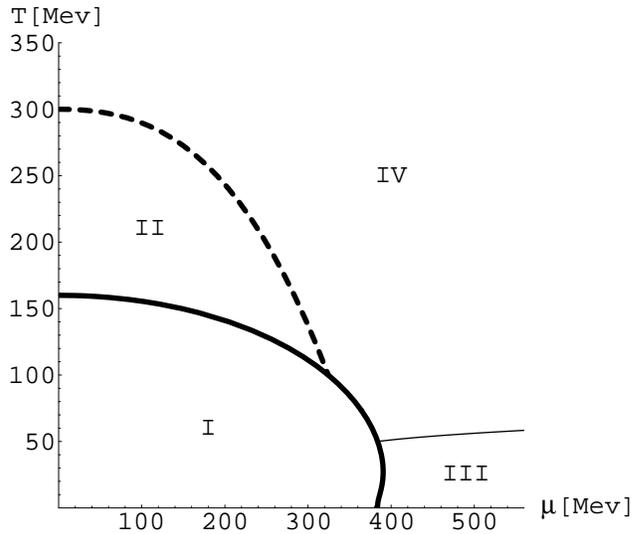}
\end{center}
\caption{QCD phase diagram: I -- hadron phase, II -- pseudogap phase,
III -- color superconducting phases, IV -- quark-gluon plasma.
}
\label{fig1}
\end{figure}

If the chiral transition is caused by pion decoherence, the
phase diagram of QCD will look as shown in Fig.~\ref{fig1}, with the pseudogap phase
sandwiched between the hadron and the quark-gluon plasma phases.
It should be mentioned that no symmetry and no order parameter can distinguish
the pseudogap phase from the quark-gluon plasma, since the separation
of the quark condensate in the phase and the modulus only makes sense
within the low-energy approximation, when phase fluctuations are sufficiently
light compared to all other modes. 
The pseudogap phase cannot be separated from quark-gluon plasma 
by a phase transition,
inasmuch as there is no phase transition between
the pseudogap state and the normal state in  \htc superconductors.
The dashed line in Fig.~\ref{fig1} thus 
denotes a smooth crossover, which can be rather broad. 
The distinctive feature of the pseudogap phenomenon is that
the symmetry restoration affects only the Goldstone modes, while parameters
associated with other excitations are continuous or almost continuous
across the phase transition. In particular,
 masses of all excitations in the pseudogap phase, except for pions, 
will be determined by constituent quark mass and thus will be rather large.
If there is no pseudogap phase and the constituent quark
mass disappears above the chiral transition, 
masses of non-Goldstone modes  are expected to
drop at the critical temperature. 

A particular mechanism that can lead to pion decoherence is
disordering of the chiral condensate by baryons \cite{DeT90,Kog01}.
The idea behind this mechanism closely follows the 
analogy with \htc superconductivity.  In the scenario proposed
in \cite{DeT90,Kog01}, baryons disorder the chiral condensate in the same way
vortices disorder the phase of the superconducting gap in two dimensions. 
The key point is that baryons can be associated with topological excitations of
the chiral field \cite{Sky62}.
Inside the baryon, 
 $U(x)$ winds around an $SU(2)$ subgroup of $SU(N_f)$. Therefore,
in a sufficiently dense random ensemble of baryons and anti-baryons, the chiral
field is randomly distributed over  $SU(N_f)$, which 
results in $\vev{\tr U}=0$. Numerical simulations of 3D sigma model  \cite{DeT90}
indicate that this picture is probably correct. In particular,
baryon susceptibility 
undergoes a dramatic rise in the vicinity of 
the phase transition \cite{DeT90}.
It was independently observed \cite{Kog01}
the thermal  density of an ideal baryon gas 
is comparatively large already 
at $T\sim 150$-$200$~Mev, despite
a small Boltzmann
factor associated with large baryon masses. The latter is
 compensated by a large entropy due to a large number of baryon 
resonances.
The phase transition in the non-linear
sigma model from this point of view resembles the BKT transition \cite{Kog01}. 
The pseudogap phase is then analogous to the high-temperature,
plasma phase of 2D XY model, in which vortices are liberated and
 all correlations are screened.
Similarly, there is no pions in the pseudogap phase and all
excitations of the chiral field must be massive. 
 
The whole idea of the pseudogap 
mechanism relies on the assumption that
 the low-energy approximation is still accurate near 
the chiral phase transition.
Therefore, the 
effective chiral Lagrangian
\rf{nlsm} should still make sense in the pseudogap phase, 
though chiral perturbation theory, at least
in its straightforward implementation, should break down.
Since the  linear and the non-linear sigma models share the same symmetries
and belong to the same universality class,
the order of the transition is determined by universality arguments of
and depends on $N_f$ \cite{Pis84}.
For realistic quark masses, Monte Carlo simulations 
indicate \cite{Kar00} that the
transition becomes sharp but smooth crossover.

The phase transition in the non-linear sigma model is a non-perturbative
phenomenon which is very hard to describe analytically. Nevertheless,
some information can be deduced from simple dimensional
arguments.
The only dimensionful parameter of the non-linear sigma model
is the pion decay constant. Consequently,
the critical temperature  
should be proportional to the pion
decay constant with some numerical coefficient:
\eq{\label{ct}
T_c\propto F_\pi.
}
Here $F_\pi$ stands for the
 ``bare'' pion decay constant, that is, the coefficient in front
of the kinetic term in the chiral Lagrangian obtained after integrating
out all heavy degrees of freedom.
Taking into account thermal
pion loops, which 
effectively reduce $F_\pi$ \cite{Gas87}, would be a double counting.
Once the 
dependence of the decay constant on the temperature and the chemical potential
is known,
the condition \rf{ct} 
can be used to locate the critical line in the $T$--$\mu$
plane:
\eq{\label{cnd}
\frac{F_\pi^2(T_c(\mu),\mu)}{T_c^2(\mu)}
=\frac{F_\pi^2(T_c(0),0)}{T_c^2(0)}.
}

I will calculate $F_\pi$ as a function of temperature and chemical 
potential in the framework of semi-phenomenological 
constituent quark model of Ref.~\cite{Dia86,Dia97},
which has been 
rather successful in describing chiral dynamics and
the nucleon properties
\cite{Dia88,Dia97}. 
Pions arise in this model as chiral phases of the constituent quark mass:
\eq{\label{qm}
{\cal L}=\bar{\psi}\br{i \d\/-M\e^{i\gamma^5\lambda^a\pi^a}}\psi
=
\bar{\psi}\left[i \d\/-M\br{
\frac{1-\gamma^5}{2}\,U+\frac{1+\gamma^5}{2}\,U\dd}\right]\psi.
}
This form of constituent quark Lagrangian was motivated
by the instanton liquid model  \cite{Dia86}. 
In fact, such an interaction of constituent quarks with pions
 will arise after Hubbard-Stratonovich 
transformation from any four-quark
interaction, local or non-local, which respects the symmetries of QCD. 
The chiral Lagrangian in this model 
is obtained after integration over quark fields and subsequent
derivative expansion of the fermion determinant.
The pion decay constant is the coefficient before the
first term with lowest number of derivatives.
The expression for pion decay constant obtained in this way
\cite{Dia88,Dia97} 
can be easily generalized to the case of
non-zero temperature and chemical potential:
\ar{\label{fpi}
F^2_\pi(T,\mu)
&=&4N_cM^2
T\sum_n\int\frac{d^3p}{(2\pi)^3}\,
\frac{1}{\left\{\left[(2n-1)\pi T-i\mu\right]^2+p^2+M^2\right\}^2}
\non
&=&F^2_\pi
-\frac{N_cM^2}{2\pi^2}
\int_M^\infty\frac{d\o}{\sqrt{\o^2-M^2}}\,
\left[\frac{1}{\e^{(\o-\mu)/{T}}+1}+\frac{1}{\e^{(\o+\mu)/{T}}+1}
\right],
}
where $N_c=3$ is the number of colors. Both the temperature and the chemical
potential tend to decrease $F_\pi$, so larger chemical potential require
lower critical temperature to satisfy Eq.~\rf{cnd}, as expected.

Lattice simulations give $T_c(0)=150$~Mev at $N_f=3$ and $T_c(0)=170$~Mev at
$N_f=2$ in the chiral limit \cite{Kar00}. The boundary of the hadron phase
 in Fig.~1 is obtained by solving Eq.~\rf{cnd}
with $T_c(0)=160$~Mev, $M=350$~Mev, and
$F_\pi=93$~Mev. Other 
lines in Fig.~1 are drawn somewhat arbitrarily. Since 
the critical temperature changes slowly in a rather wide range of
chemical potentials and is always smaller than the constituent quark mass,
Boltzmann statistics should be a good approximation unless $\mu$ is
close to $M$. The expansion in $T/M$ then yields the 
following analytic expression for
the critical line at 
$\Delta T\equiv T_c(\mu)-T_c(0)\ll T_c(0)$:
\eq{
\frac{\Delta T}{T_0}\approx \frac{N_cM^2}{2\pi F_\pi^2}\,
\sqrt{\frac{T_0}{2\pi M}}
\,\e^{-M/{T_0}}\br{\cosh\frac{\mu}{T_0}-1},
}
where $T_0\equiv T_c(0)$.

When $T$ is small and $\mu$ is sufficiently large, the transition to the
color superconducting state is expected to occur \cite{Raj00}.
This transition cannot be driven by pion disordering, so the condition \rf{cnd} 
is expected to
work only for $\mu<M$. Nevertheless, the equation $F_\pi(0,\mu_c)=0$, to which
\rf{cnd} reduces at zero temperature, gives a reasonable value of $\mu_c$
which is roughly consistent with various estimates 
 of a critical chemical potential for
color superconducting phase transition \cite{Ber99}.

The chiral symmetry is broken in a color
superconductor by the diquark condensate whose
phase excitations are similar to ordinary pions \cite{Alf99} and are massless
in the chiral limit. However, the pion decay constant in a color 
superconductor 
is rather large \cite{Son00}
and it is very unlikely that pion decoherence can
emerge in the color superconducting phase.
 
Finally, I should mention that the
pseudogap phase, if exists at $N_c=3$, disappears in the large-$N_c$
limit\footnote{ Refs.~\cite{Kle98,Bab01} contain a detailed discussion of what happens to the
pseudogap phase in the large-$N$ limit of $(2+\varepsilon)$-dimensional
Gross-Neveu model}. 
The pion decay constant (chiral phase stiffness) grows
as $F_\pi^2=O(N_c)$
at large $N_c$ and suppresses fluctuations
of the chiral field that could drive pion decoherence.
The condition \rf{ct} then gives $T_c=O(\sqrt{N_c})$, which is smaller
than an estimate based on the ideal gas approximation for baryons \cite{Kog01}:
$T_c=O(N_c/\ln N_c)$, but still grows with $N_c$, unlike
the temperature at which chiral
condensate completely melts, which is supposed to be finite at 
$N_c\rightarrow\infty$.

I am grateful to M.~Franz for
very helpful discussions of the pseudogap phase in high-$T_c$
superconductors and to E.~Babaev and
A.~Zhitnitsky for interesting comments.
This work was supported by 
NSERC of Canada, by Pacific Institute for the Mathematical Sciences
and in part by RFBR 
 grant 01-01-00549 and RFBR grant
00-15-96557 for the promotion of scientific schools.


\end{document}